\documentclass[12pt,amsfonts]{article}

\usepackage{amsmath}

\usepackage{amssymb}
\usepackage{bm}
\usepackage{graphicx}
\usepackage{xcolor}
\usepackage{caption}
\usepackage{subcaption}
\usepackage{appendix}
\usepackage{bigints}
\usepackage{yfonts}
\usepackage{bbold}
\usepackage[T1]{fontenc} % if needed
\usepackage{amssymb}
\usepackage{bm}
\usepackage{graphicx}
\usepackage{xcolor}
\usepackage{authblk}
\usepackage{hyperref}

\newcommand{\be}{\begin{equation}}
\newcommand{\ee}{\end{equation}}
\newcommand{\ba}{\begin{array}}
\newcommand{\ea}{\end{array}}
\newcommand{\bqa}{\begin{eqnarray}}
\newcommand{\eqa}{\end{eqnarray}}
\newcommand{\bea}{\begin{eqnarray}}
\newcommand{\eea}{\end{eqnarray}}

\begin{document}

\newtheorem{defi}{Definition}[section]
\newtheorem{lem}[defi]{Lemma}
\newtheorem{prop}[defi]{Proposition}
\newtheorem{theo}[defi]{Theorem}
\newtheorem{rem}[defi]{Remark}
\newtheorem{cor}[defi]{Corollary}

\title{Quasi-integrability from ${\cal PT}$-symmetry}

\author[1]{Kumar Abhinav\thanks{kumar.abh@mahidol.ac.th}}
\author[2]{Partha Guha\thanks{partha.guha@ku.ac.ae}}
\author[3]{Indranil Mukherjee\thanks{indranil.mukherjee@makautwb.ac.i}}
\affil[1]{Centre for Theoretical Physics and Natural Philosophy, Nakhonsawan Studiorum for Advanced Studies, Mahidol University, Nakhonsawan 60130, Thailand}
\affil[2]{Department of Mathematics, Khalifa University of Science and Technology, PO Box 127788, Abu Dhabi, UAE}
\affil[3]{School of Management Sciences, Maulana Abul Kalam Azad University of Technology, Haringhata, Nadia, Pin-741249, India}

\date{\today}
\maketitle
\abstract{Parity and time-reversal (${\cal PT}$) symmetry is shown as the natural cause of quasi-integrability of deformed integrable models, crucial to represent real physical systems as they posses various irregularities. The condition for asymptotic conservation of quasi-conserved charges appear as a direct consequence of the ${\cal PT}$-symmetric phase of the system, ensuring definite ${\cal PT}$-properties of the corresponding Lax pair as well as that of the anomalous contribution, consistent with the Wilson-loop criterion for integrability-like behavior. As a result, the quasi-deformed charge densities always acquire definite ${\cal PT}$-properties suitable for the asymptotic conservation, as the Abelianization approach to construct them also preserves the definite ${\cal PT}$-behavior of the Lax pair. This ${\cal PT}$-symmetry based origin of quasi-conservation is general and has been demonstrated for quasi-deformations of multiple systems such as KdV, NLSE and non-local NLSE.}



\section{Introduction}
Integrable continuous and thus infinite-dimensional models are characterized by an infinity of conserved N\"other charges \cite{R26}. They represent an infinite number of continuous symmetries, each characterized by an equation of continuity of the form $\rho_t(x,t)+j_x(x,t)=0$, where $\rho(x,t)$ and $j(x,t)$ depend on the dynamical variable of the system and the suffixes represent partial derivatives, with the corresponding conserved N\"other charge being,
\begin{equation}
    Q=\int\,dx\,\rho(x,t),\quad\frac{dQ}{dt}=0.\label{NEq7}
\end{equation} 
These symmetries can support stable localized excitations like solitons and kinks etc which are robust against the usual dispersion owing to the system's dynamics. Formally, the Lax formalism \cite{Lax} ensures the integrability of a continuous system by morphing its governing differential equation into the zero curvature condition \cite{AKNS},
\begin{equation}
    F_{tx}=A_t-B_x+\left[A,\,B\right]=0,\label{NEq8}
\end{equation}
through the Lax pair $(A,B)$. However, real continuous systems such as oceans, atmospheric flows, and biological fluids are known to support similar localized structures, which are fairly stable against dispersions. For example, solitary waves have been physically observed in fluids and kinks appear in various physical systems \cite{SolHis} despite the lack of infinite continuous symmetries due to physical irregularities, impurities, deformations, etc. Consequently, for realistic modeling of continuous physical systems, quasi-integrable deformations (QID) of integrable models \cite{R1,R2} appear to be the natural choice. In such systems, each degree of freedom does not correspond to a continuous symmetry, and thus an infinite subset of charges becomes {\it anomalous}. Still, they possess localized single- and multi-soliton and kink-like structures \cite{R1,R2} which is attributed to the property that although these non-conserved charges vary locally, they approach fixed values at the spatio-temporal infinities $(x\to\pm\infty,t\to\pm\infty)$. For this to happen, it is required that the densities $\rho_i(x,t)$s corresponding to these {\it anomalous} charges have suitable definite behavior (even or odd) under parity (${\cal P}: x\to -x$) and time-reversal (${\cal T}: t\to -t$) \cite{R3}. More specifically, the integrand in the expression,
\begin{equation}
    \frac{dQ}{dt}=\int\,dx\,\rho_t(x,t),\label{NEq9}
\end{equation}
must be odd under ${\cal PT}$ to yield,
\begin{equation}
    \int_{-T}^T\,dt\,\frac{dQ}{dt}=Q(t=T)-Q(t=-T)=0,\quad T\to\infty.\label{NEq10}
\end{equation}
In the last few years, quasi-deformations of various integrable models such as sine-Gordon \cite{R1} and its supersymmetric counterpart \cite{R4}, nonlinear Schr\"odinger \cite{R3} and related hierarchies \cite{R5}, AB \cite{R6} and KdV \cite{R7,R8} etc have been obtained mainly through loop-algebra-based Abelianization \cite{R2} and also via a Riccati-type pseudo-potential approach \cite{R8,R9}. There, single and multiple soliton-like structures emerge that are fairly stable and robust under scattering \cite{R1,R2,R3,R8,R9}.

\paragraph*{}Quasi-deformation is achieved by modifying the temporal Lax component ($B$) of an integrable system by deforming the corresponding potential \cite{R1,R2} or Hamiltonian \cite{R7}. The deformed equation then can be justifiably expanded in the deformation parameter $\epsilon$ \cite{R3}, leading to a form,
\begin{equation}
    F^d_{tx}={\cal X}_xM,\label{Eq1}
\end{equation}
where the deformed curvature $F^d_{tx}=A_{d,t}-B_{d,x}+\left[A_d,\,B_d\right]$ is made of the deformed Lax pair $\left(A_d,B_d\right)$ having an inherent Lie algebra basis to which the operator $M$ belongs. The deviation from the zero-curvature (integrability) condition is caused by the anomaly function ${\cal X}$ which is ${\cal O}(\epsilon)$. The Abelianization approach \cite{R1,R2} then most naturally leads to anomalous charges that satisfy,
\begin{equation}
    \frac{dQ_d^n}{dt}=\int dx\,{\cal X}f_n\left(u_d\right)\neq 0,\label{Eq2}
\end{equation}
where $f_n$ are some functions of the deformed solution $u_d(x,t)$. Naturally, the asymptotic charge difference $Q^n(t=\infty)-Q^n(t=-\infty)$ can vanish if the factor ${\cal X}f_n$ is odd in both parity (${\cal P}$) and time-reversal (${\cal T}$) for a given deformed solution.

\paragraph*{} The ${\cal PT}$-dependent behavior for asymptotic conservation laws follows from how the system behaves under the same transformations. Although invariance under both these transformations is expected from almost all physical systems, the complete consequences of the same are seldom utilized in the presence of the self-adjoint (Hermitian) nature of the system. This is particularly true in the case of linear continuous systems, which are usually described in a complex vector space with square-integrability. Subsequently, non-Hermitian ${\cal PT}$-symmetric linear systems have garnered wide interest because they possess real eigenvalues in a particular (symmetric) parametric phase and complex-conjugate pairs otherwise (broken phase) \cite{R10,R11,R12}. 
The symmetric phase is intriguing, since in a square-integrable complex vector space, only Hermitian operators are ensured of real eigenvalues:
\begin{equation}
    H\vert\psi\rangle=E\vert\psi\rangle\Rightarrow\langle\psi\vert H\vert\psi\rangle-\langle\psi\vert H^\dagger\vert\psi\rangle=\left(E-E^*\right)\langle\psi\vert\psi\rangle.\label{NEq11}
\end{equation}
This eventually mandates a pseudo-Hermitian extension \cite{Most} of the normed vector space for ${\cal PT}$-symmetric non-Hermitian systems as,
\begin{equation}
\langle\psi\vert\psi\rangle\to\langle\psi\vert\eta\vert\psi\rangle\Rightarrow\langle\psi\vert H\vert\psi\rangle\to\langle\psi\vert H\eta\vert\psi\rangle,\quad H^\dagger\eta=\eta H,\label{NEq12}
\end{equation}
with the {\it metric operator} $\eta$. The transition between symmetric and broken phases (real and complex-conjugate spectrum) is a proper phase transition due to the spontaneous breaking of the ${\cal PT}$-symmetry, with the coalescence of eigenstates from the broken to the symmetric phase, the latter showing definite ${\cal PT}$-properties of the eigenstates \cite{R13,R10,R12}. Non-Hermitian ${\cal PT}$-symmetric analogs have also been studied for nonlinear \cite{R14} and field theoretical \cite{R15,R16,R17} systems that have generalized the relation between symmetries and conservations \cite{R17}. In particular, for nonlinear systems, the ${\cal PT}$ eigenvalues of symmetric solutions are restricted by nonlinearity \cite{R18,R19,R20}.  

\paragraph*{} The phenomena of possessing localized solutions despite the lack of integrability and presence of real eigenvalues in absence of hermiticity have a similar tone, more so since definite ${\cal PT}$-property is necessary for quasi-conservation. This connection between ${\cal PT}$-symmetry and QID was first attempted by Assis \cite{R21} in a Wilson loop approach \cite{R22} in which the presence of ${\cal PT}$-symmetry mimicked integrability even in systems which did not possess a vanishing curvature a priori. This is interesting as the deforming extensions, or quasi-deformation anomalies, are usually not restricted by any such symmetry condition \cite{R1,R2}. In practice, the quasi-deformation anomaly can disrupt even the existing Hermitian structure. On the other hand, since quasi-integrability is achieved only if the final form of the quasi-conserved charges has the required ${\cal PT}$-property, it is very likely that ${\cal PT}$-symmetry needs to prevail in the deformed system. However, to our knowledge, a direct demonstration that the inherent ${\cal PT}$-symmetry of a deformed integrable system is responsible for its quasi-integrability has not yet been obtained. 

\paragraph*{}In this work, we obtain a direct connection between the ${\cal PT}$-symmetry of the system and its quasi-integrability. To this end, we found that the definite space-time parity of the deformed solution directly implies ${\cal PT}$-oddness of the anomalous integrands in time-variations of all the quasi-conserved charges. This ensures that if the deformation of an integrable system is ${\cal PT}$-symmetric then it can be quasi-integrable, in the same sense that non-Hermitian systems can possess real spectra for being ${\cal PT}$-symmetric. We demonstrate this mechanism in known quasi-deformed cases of KdV \cite{R7} and NLS \cite{R3}, together with the non-local NLS system \cite{R23} which is inherently non-hermitian yet integrable.   

\paragraph*{}In the following definite ${\cal PT}$-structure of conserved charges is demonstrated as the direct consequence of ${\cal PT}$-symmetry of the solution in section \ref{Se1}. We consider the {\bf KdV system} as the main working example for the treatment. The ${\cal PT}$-structure of the quasi-deformed counterpart is demonstrated in section \ref{Se2}. The cases of {\bf quasi-NLSE} and {\bf quasi-non-local NLSE} are dealt with later in this section, followed by demonstrating that the Abelianization procedure \cite{R1} also respects this ${\cal PT}$-QID correspondence. We conclude in section \ref{Se3} after mentioning immediate possibilities.

\section{${\cal PT}$-structure of integrability and conserved charges}\label{Se1}
The standard lore of ${\cal PT}$-symmetry is based on linear systems which are {\it not Hermitian} \cite{R10,R11,R12}. It is the ${\cal PT}$-symmetric phase, hosting real spectra in the absence of Hermiticity along with ${\cal PT}$-definite eigenstates, that mandates a generalization of the Hilbert space akin to the pseudo-Hermitian systems \cite{R24}. For their nonlinear counterparts, the analogous identifier of the symmetric phase is a definite-${\cal PT}$ solution. Given that the system is also integrable, the zero curvature condition $F_{tx}=0$ must also survive the ${\cal PT}$-transformation as,
\begin{equation}
    F^{PT}_{tx}=-A^{PT}_t+B^{PT}_x+\left[A^{PT},\,B^{PT}\right]=0.\label{Eq3}
\end{equation}
This implies a new Lax pair $L^{PT}_\mu=\left(-A^{PT},-B^{PT}\right)$ that must identify with the original pair $L_\mu=(A,B)$ as,
\begin{eqnarray}
    &&\quad \left[L_\mu\left(u,u_t,u_x,u_{xt},\cdots\right)\right]^{PT}=-L_\mu\left(u^{PT},u^{PT}_t,u^{PT}_x,u^{PT}_{xt},\cdots\right)\nonumber\\
   && \Rightarrow L_\mu^*\left(u^{PT},-u^{PT}_t,-u^{PT}_x,u^{PT}_{xt},\cdots\right)=-L_\mu\left(u^{PT},u^{PT}_t,u^{PT}_x,u^{PT}_{xt},\cdots\right)\nonumber\\
   && \Rightarrow L_\mu^*\left(u,-u_t,-u_x,u_{xt},\cdots\right)=-L_\mu\left(u,u_t,u_x,u_{xt},\cdots\right).\label{Eq4}
\end{eqnarray}
for the system's ${\cal PT}$-symmetry, {\it i. e.}, resulting in the same equation with solution $u^{PT}=u^*(-x,-t)$. Therefore, for a ${\cal PT}$-symmetric integrable model, the functional form of the Lax pair is ${\cal PT}$-odd. The last equation is the result of another ${\cal PT}$-operation, which is also true if $u^{PT}=u$. Notably, the ${\cal PT}$-oddness of the Lax pair is a consequence of a non-trivial commutator in the curvature, attributed to nonlinearity of the system which in turn determines the sign of $u$ under ${\cal PT}$-operation in the symmetric phase. Consequently, the other choice for the unbroken phase $u^{PT}=-u$, instead leads to,
\begin{equation}
    L_\mu^*\left(-u,u_t,u_x,-u_{xt},\cdots\right)=-L_\mu\left(-u,-u_t,-u_x,-u_{xt},\cdots\right).\label{Eq5}
\end{equation}
Eq. \ref{Eq4} is always valid for a given ${\cal PT}$-symmetric system, whereas Eq. \ref{Eq5} may not be allowed simultaneously, especially when the system is nonlinear\footnote{For ${\cal PT}$-symmetric nonlinear models, it is the nonlinear term that picks the sign of the symmetric phase $u^{PT}=\pm u$.}.  

\paragraph*{}As a definite case, consider the {\bf KdV equation} that maintains its form under ${\cal PT}$-operation as,
\begin{equation}
    u_t=uu_x+u_{xxx}\Rightarrow u^{PT}_t=u^{PT}u^{PT}_x+u^{PT}_{xxx},\quad u^{PT}=u^*(-x,-t).\label{E25}
\end{equation}
The unbroken phase exclusively corresponds to $u^{PT}=u$ given the particular nonlinearity, simplifying the condition for the Lax pair as,
\begin{equation}
    L_\mu^{PT}(x,t)=L_\mu^*(-x,-t)=-L_\mu(x,t).\label{Eq6}
\end{equation}
It is a combined effect of the ${\cal PT}$-symmetry and the particular unbroken sector ($u^{PT}=u$). Indeed, the explicit KdV Lax pair,
\begin{equation}
    A=\sigma_+-\frac{u}{6}\sigma_-,\quad B=\frac{1}{6}\left(u_x\sigma_3-2u\sigma_++\left(u_{xx}+\frac{u^2}{3}\right)\sigma_-\right),\label{Eq6a}
\end{equation}
in the usual $su(2)$ basis, is ${\cal PT}$-odd in the unbroken phase. The Pauli matrices have characteristic ${\cal PT}$-properties,
\begin{equation}
    \sigma_\pm^{PT}:={\cal PT}\sigma_\pm\left({\cal PT}\right)^{-1}=-\sigma_\pm,\quad\sigma_3^{PT}:={\cal PT}\sigma_3\left({\cal PT}\right)^{-1}=\sigma_3,\label{E56}
\end{equation}
under the identifications,
\begin{equation}
    {\cal P}=\sigma_x\quad\text{and}\quad{\cal T}=\sigma_yK,\quad K: i\to -i.\label{E55}
\end{equation}

\paragraph*{}Eventually, all the conserved KdV charges correspond to ${\cal PT}$-even densities (integrands)\footnote{This makes sense as only then their time derivatives will contain ${\cal PT}$-odd integrands, thereby leading to conservation.}. The first few of them are listed below,  
\begin{eqnarray}
     Q^1(t)&=&\int\,dx\,u(x,t),\nonumber\\
     Q^2(t)&=&\int\,dx\,\frac{u^2(x,t)}{2},\nonumber\\
     Q^3(t)&=&\int\,dx\,\left(\frac{u^3}{3}-\frac{u_x^2}{2}\right),\nonumber\\
     Q^4(t)&=&\frac{1}{6}\int\,dx\,\left(\frac{5}{12}u^4-5uu_x^2+3u_{xx}^2\right),\nonumber\\
     \vdots,\label{Eq7}
\end{eqnarray}
which are conserved in the usual way via the KdV equation \cite{R25}. In general, consider the undeformed charge,
\begin{equation}
    Q^{n+1}=\int\,dx\,\rho^{n+1},\label{E51}
\end{equation}
wherein the densities $\rho^{n+1}=3(-1)^nv_{2n}$ follow the recursion relation \cite{R26},
\begin{equation}
    v_n=-iv_{n-1,x}-\frac{1}{6}\sum_m^{n-2}v_{n-2-m}v_m.\label{E45}
\end{equation}
Upon taking a time-derivative one gets,
\begin{eqnarray}
    \frac{dQ^{n+1}}{dt}&=&\int\,dx\,\sum_k^{2n-2}\frac{\partial\rho^{n+1}}{\partial u^{(k)}}u^{(k)}_t,\quad u^{(k)}=\frac{\partial^ku}{\partial x^k}\nonumber\\
    &=& \sum_k^{2n-2}\int\,dx\,(-1)^k\left(\frac{\partial\rho^{n+1}}{\partial u^{(k)}}\right)^{(k)}u_t.\label{E52}
\end{eqnarray}
The order of $x$-derivative $k$ goes up to $2n-2$ for $v_{2n}$ \cite{R25,R26}. The last integral can vanish, implying integrability, if the term multiplying $u_t$ is ${\cal PT}$-even {\it i. e.},
\begin{equation}
    (-1)^{2k}\left(\frac{\partial\rho_{PT}^{n+1}}{\partial u^{(k)}}\right)^{(k)}=\left(\frac{\partial\rho^{n+1}}{\partial u^{(k)}}\right)^{(k)}\Rightarrow\rho_{PT}^{n+1}=\rho^{n+1}.\label{E53}
\end{equation}
Therefore, the integrand in the expression for $dQ_d^{n+1}/dt$ will always be ${\cal PT}$-odd in the unbroken phase ($u^{PT}=u$). A more direct route to the ${\cal PT}$-behavior of the densities is through substitution. Starting with $v_0=u$ and then on order-by-order substitution in the recursion relation yields,
\begin{eqnarray}
    &&\rho^1=u,\nonumber\\
    &&\rho^2=\frac{u^2}{2}+\text{T.D.},\nonumber\\
    &&\rho^3=\frac{1}{6}u^3+4u_x^2+uu_{xx}+\text{T.D.},\nonumber\\
    &&\vdots,\label{E53A}
\end{eqnarray}
where T.D. stands for total derivative. From Eq. \ref{E45}, every $v_n$ leads to $v_{n-1}$ by a single derivative, and the bilinear terms are always in even-even or odd-odd pairing. Thus, every $v_{2n}$ that contributes to $\rho^{n+1}$ will contain an even order of derivatives (modulo T.D.s), which implies $\rho_{PT}^{n+1}=\rho^{n+1}$ in the unbroken phase. Since the definite ${\cal PT}$-behavior of the solution is synonymous with integrability, the broken phase ($u^{PT}\neq u$) may not be observed in a ${\cal PT}$-symmetric integrable system.

\section{${\cal PT}$-structure of quasi-conservation}\label{Se2}
Upon quasi-deformation, the previously integrable equation develops an anomaly ${\cal Y}={\cal X}_x$ as,
\begin{equation}
    F_{d,tx}=A_{d,t}-B_{d,x}+\left[A_d,\,B_d\right]={\cal Y}M,\label{E32}
\end{equation}
where the suffix $d$ signifies quasi-deformation and $M$ belongs to the governing algebra of the integrable counterpart. ${\cal Y}$  is at least first order in the deformation parameter $\varepsilon$ \cite{R1,R3}. Under ${\cal PT}$, this nonzero curvature changes as,
\begin{equation}
    F^{PT}_{d,tx}=-A^{PT}_{d,t}+B^{PT}_{d,x}+\left[A^{PT}_d,\,B^{PT}_d\right]={\cal Y}^{PT}M^{PT},\label{Eq8}
\end{equation}
mandating a ${\cal PT}$-odd Lax pair $L_\mu^d=\left(A_d,B_d\right)$ again for overall ${\cal PT}$-symmetry, along with a ${\cal PT}$-even product ${\cal Y}^{PT}M^{PT}$. This result seems to contradict the assertion in Ref. \cite{R21}, in which the phase of the Wilson loop operator is responsible for the evolution of the system \cite{R22},
\begin{equation}
    W\left(\Gamma\right)={\cal P}_\Gamma\exp\left[-\oint_\Gamma L_\mu dx^\mu\right]\equiv{\cal P}_\Gamma\exp\left[-\int_\Gamma\,\left(Adx+Bdt\right)\right],\label{E27A}
\end{equation}
with path-ordering ${\cal P}_\Gamma$ over a closed loop $\Gamma$, vanished non-trivially for ${\cal PT}$-{\it even} $L_\mu=\left(A,\,B\right)$. This operator governs the phase of evolution of a field $\Psi$ over the gauge group that characterizes $L^d_\mu$. The non-trivial vanishing of the phase, when the integrand is non-zero, is akin to quasi-integrability \cite{R21} following the correspondence \cite{R22},
\begin{eqnarray}
    {\cal P}_\Gamma\exp\left(-\oint_\Gamma\,d\sigma\,A_\mu\frac{dx^\mu}{d\sigma}\right)&=&{\cal P}_\Sigma\exp\left(-\int_\Sigma\,d\sigma\,d\tau\,\Psi^{-1}F_{\mu\nu}\Psi\frac{dx^\mu}{d\sigma}\frac{dx^\nu}{d\tau}\right),\label{E27B}
\end{eqnarray}
where ${\cal P}_\Sigma$ signifies path-ordering for the integral over the area $\Sigma$ enclosed by $\Gamma$ parameterized by the parameters $\sigma$ and $\tau$ respectively. As a particular example, considering a square loop $\left\{(-L,-\tau),\,(L,-\tau),\,(L,\tau),\,(-L,\tau)\right\}$ in the $(x,t)$-plane, 
\begin{eqnarray}
    \oint_\Gamma L_\mu dx^\mu&\equiv&\int_{-L}^L\left(A^{PT}(x,\tau)-A(x,\tau)\right)dx+\int_{-\tau}^\tau\left(B(L,t)-B^{PT}(L,t)\right)dt,\label{E77}
\end{eqnarray}
that clearly vanish for a ${\cal PT}$-even Lax pair \cite{R1}. However, since the intervals for both integrals are symmetric ${\cal PT}$-oddness of the above integrands will also serve the purpose. This is clearer from the re-arrangement,
\begin{eqnarray}
    \oint_\Gamma L_\mu dx^\mu&\equiv&\int_0^L\left[\left\{A(x,-\tau)+A^{PT}(x,-\tau)\right\}-\left\{A(x,\tau)+A^{PT}(x,\tau)\right\}\right]dx\nonumber\\
    &+&\int_0^\tau\left[\left\{B(L,t)+B^{PT}(L,t)\right\}-\left\{B(-L,t)+B^{PT}(-L,t)\right\}\right]dt,\label{E80}
\end{eqnarray}
as all four braces above vanish for a ${\cal PT}$-odd Lax pair. This situation is more favorable, as it follows from the imposed ${\cal PT}$-symmetry of the system that will be shown to yield quasi-conserved charges. 

\subsection{A ${\cal PT}$-induced quasi-KdV system}
As a particular case, the {\bf quasi-KdV equation} \cite{R7} now transforms under ${\cal PT}$ to, 
\begin{equation}
    u_{d,t}=u_du_{d,x}+u_{d,xxx}+{\cal Y}\Rightarrow u^{PT}_{d,t}=u^{PT}_du^{PT}_{d,x}+u^{PT}_{d,xxx}-{\cal Y}^{PT}.\label{E33}
\end{equation}
Demanding ${\cal PT}$-symmetry now with the symmetric phase $u^{PT}_d=u_d$ invariably leads to,
\begin{equation}
    {\cal Y}^{PT}(x,t)={\cal Y}^*(-x,-t)=-{\cal Y}(x,t).\label{E35}
\end{equation}
This would directly imply $M^{PT}=-M$ from Eq. \ref{Eq8}, with the ${\cal PT}$-oddness of the quasi-deformed Lax pair $\left(A_d,B_d\right)$ carried over from the undeformed case\footnote{The quasi-deformed KdV Lax pair is provided in Ref. \cite{R7} which follows this assertion.}. As for the quasi-conserved charges, the $u\to u_d$ analogs of those given in Eq. \ref{Eq7} serves the purpose nicely in the sense that they should be conserved for ${\cal Y}=0$\footnote{On taking the time-derivative of these charges, the integral will always be linear in $u_{d,t}$, which when replaced from the quasi-KdV equation the only non-vanishing (non-T.D.) contribution will be linear in ${\cal Y}$.}. The first few anomalous conservation laws for this deformed KdV system are,
\begin{eqnarray}
    &&\frac{d}{dt}Q^1_d=\int\,dx\,{\cal Y},\nonumber\\
    &&\frac{d}{dt}Q^2_d=\int\,dx\,u_d{\cal Y},\nonumber\\
    &&\frac{d}{dt}Q^3_d=\int\,dx\,\left(\frac{u_d^2}{2}+u_{d,xx}\right){\cal Y},\nonumber\\
    &&\frac{dQ_d^4}{dt}=\frac{1}{6}\int\,dx\,\left(\frac{5}{3}u_d^3+5u_{d,x}^2+10uu_{d,xx}+6u_{d,xxxx}\right){\cal Y},\nonumber\\
    &&\vdots\label{Eq9}
\end{eqnarray}
All of the above integrands are ${\cal PT}$-odd in the symmetric phase, leading to quasi-conservation. This is a general result coming from the ${\cal PT}$-even deformed densities $\rho^{n+1}_d$ corresponding to ($u\to u_d$) in Eq.s \ref{Eq7} as the time-derivative only lowers the power of $u_d$ therein by one. In particular, the required even-${\cal PT}$ condition is fulfilled by the recently reported single-soliton solution for a particular quasi-deformed KdV system \cite{R026,R026a},
\begin{equation}
    u^{\text{1-Sol}}_d(x,t)=\frac{k^2}{(2+\varepsilon_2)\left(1-\varepsilon_1k^2\right)}\text{sech}^2\left[\frac{1}{2}(kx-\omega t+\delta)\right],\label{NEq1}
\end{equation}
wherein $\varepsilon_{1,2}$ are the deformation parameters with wave-number $k$, frequency $\omega$ and constant phase $\delta$. Furthermore, a 2-soliton ${\cal PT}$-even quasi-KdV structure exists for $\varepsilon_{1,2}=1$ which has the form \cite{R026a},
\begin{equation}
    u^{\text{2-Sol}}_d(x,t)=-\partial_x\partial_t\log\left[e^{\Delta/4}\cosh\left(\frac{\eta_1+\eta_2}{2}\right)+e^{-\Delta/4}\cosh\left(\frac{\eta_1-\eta_2}{2}\right)\right],\label{NEq2}
\end{equation}
where $\eta_{1,2}=k_{1,2}x-\omega_{1,2}t+\delta_{1,2}-\Delta/2$ with $\Delta$ being another parameter. Additional numerical solutions \cite{R026,R026a} involving the multi-soliton interaction \cite{R026b} further provide ${\cal PT}$-even solutions for quasi-KdV systems.

\subsection{${\cal PT}$-induced quasi-conservation in other models} 
A similar ${\cal PT}$-based quasi-conservation structure can also be observed in other quasi-deformed models. In particular, we consider quasi-deformations of the NLS equation \cite{R3} and its ${\cal PT}$-symmetric non-local version \cite{R23} for demonstration. The {\bf quasi-NLS model} undergoes ${\cal PT}$-transformation as,
\begin{eqnarray}
    iq_{d,t}&=&-\frac{1}{2}q_{d,xx}+\kappa\vert q_d\vert^2q_d+{\cal Y}\nonumber\\
    \Rightarrow iq^{PT}_{d,t}&=&-\frac{1}{2}q^{PT}_{d,xx}+\kappa\left\vert q^{PT}_d\right\vert^2q^{PT}_d+{\cal Y}^{PT},\label{E66}
\end{eqnarray}
requiring ${\cal Y}^{PT}={\cal Y}$ for ${\cal PT}$-symmetry in the non-unique ${\cal PT}$-symmetric phase $q_d^{PT}=q_d$. In addition, the combination ${\cal Y}^{PT}=-{\cal Y}$ with $q_d^{PT}=-q_d$ also yields a symmetric phase, indicating a possible degeneracy in the spectrum. The quasi-conserved charges can be borrowed from the undeformed system as,
\begin{equation}
    Q_d^1=\int\,dx\,\vert q_d\vert^2,\quad Q^2=\int\,dx\,\left(q^*_dq_{d,x}-q_dq^*_{d,x}\right),\quad Q^3=\frac{1}{2}\int\,dx\,\left(\vert q_{d,x}\vert^2+\kappa\vert q_d\vert^4\right),\cdots\label{E68}
\end{equation}
These charges are generated through the recursion relation among the respective densities \cite{R27},
\begin{equation}
    \rho^{n+1}_d=\frac{1}{2}\left(q_d^*\left(\frac{\rho_d^n}{q_d^*}\right)_x-q_{d,x}^*\frac{\rho_d^n}{q_d^*}\right)-\kappa\sum_{i+j=n-1}\rho_d^i\rho_d^j.
\end{equation}
with $\rho_d^0=\vert q_d\vert^2$. Since $\rho_d^0$ is both real and ${\cal PT}$-even, $\rho_d^1$ is a constant ($i$) times the imaginary part of a ${\cal PT}$-odd function which has to be ${\cal PT}$-even. Subsequently, $\rho_d^2$ will also be proportional to a ${\cal PT}$-even function and by succession,
\begin{equation}
    \rho^n_d\sim{\rm Const}.\times\Psi_E,
\end{equation}
where $\Psi_E$ are ${\cal PT}$-even functions. Now in the time variation of a general deformed charge,
\begin{eqnarray}
    \frac{dQ_d^n}{dt}=\int\,dx\left[\sum_k(-1)^k\left(\frac{
    \partial\rho^n_d}{\partial q_d^{(k)}}\right)^{(k)}(-i{\cal Y})+\sum_k(-1)^k\left(\frac{
    \partial\rho^n_d}{\partial q_d^{*,(k)}}\right)^{(k)}(i{\cal Y}^*)\right].\label{E029}
\end{eqnarray}
the integrand is just some constant ($i$) multiplying the imaginary part of a ${\cal PT}$-even operator (which includes ${\cal Y}$) acting on $\rho^n_d$. Therefore, the whole integrand is proportional to a ${\cal PT}$-odd function as needed for quasi-conservation. In particular, the time variations of the charges in Eqs. \ref{E66},
\begin{eqnarray}
 &&\frac{dQ^1_d}{dt}=2\int\,dx\,\text{Im}\left(q_d^*{\cal Y}\right),\nonumber\\
    &&\frac{dQ^2_d}{dt}=4i\int\,dx\,\text{Re}\left(q_{d,x}{\cal Y}^*\right),\nonumber\\
    &&\frac{dQ^3_d}{dt}=-2\int\,dx\,\text{Re}\left(q_{d,t}^*{\cal  Y}\right).\label{E70}
\end{eqnarray}
all have ${\cal PT}$-odd integrands as required. To see this, in particular, consider the product $q_d^*{\cal Y}$ which is always ${\cal PT}$-even for the present system. Obviously, its individual real and imaginary parts have to be even and odd under ${\cal PT}$ respectively. In the presence of an extra derivative, the reverse is true. Therefore, all integrands in Eq. \ref{E70} are ${\cal PT}$-odd.

\paragraph*{}A particular 1-soliton solution for quasi-NLS systems had been obtained as \cite{R3},
\begin{eqnarray}
    \left\vert q^{\text{1-Sol}}_d(x,t)\right\vert=\left[\frac{2+\varepsilon}{2}\frac{\rho^2}{\vert\eta\vert}\text{sech}^2\left\{(1+\varepsilon)(x-vt)\right\}\right]^{\frac{1}{2+2\varepsilon}},\label{NEq3}
\end{eqnarray}
where $\varepsilon$ is the quasi-deformation parameter, $\rho$ is the wave-number, and $v$ is the velocity, along with the 2-soliton structure. All of these structures are found to have definite ${\cal PT}$-properties suitable for quasi-integrability.

\paragraph*{}Another interesting system is the ${\cal PT}$-symmetric {\bf non-local NLS system} \cite{R23}, which is also integrable but non-Hermitian, having a quasi-deformed version,
\begin{equation}
    iq_{d,t}(x,t)=q_{d,xx}(x,t)+\sigma q_d(x,t)q_d^*(-x,t)q_d(x,t)+{\cal Y},\quad\sigma=\pm.\label{E73}
\end{equation}
Such non-local ${\cal PT}$-symmetric equations model real situations and have recently been found to incorporate various physically relevant solutions such as bright and dark multi-solitons as well as breathers \cite{R028}. In particular, various non-local ${\cal PT}$-symmetric extensions of the NLS system support multi-component rogue waves \cite{R029}, breathers, and bounded solitons \cite{R030} that are unique from their local counterparts. The ${\cal PT}$-symmetry of the non-local NLS system above requires ${\cal Y}^{PT}={\cal Y}$ with a symmetric phase for $q_d^{PT}=q_d$. The corresponding quasi-conserved charges,
\begin{eqnarray}
    &&Q_d^1=\int\,dx\,q_d^*(-x,t)q_d(x,t),\nonumber\\
    &&Q_d^2=\int\,dx\,\left(q_{d,x}(x,t)q_d^*(-x,t)+q_d(x,t)q_{d,x}^*(-x,t)\right),\nonumber\\
    &&Q_d^3=\int\,dx\,\left(q_{d,x}(x,t)q_{d,x}^*(-x,t)-\sigma q_d^2(x,t)q_d^{*2}(-x,t)\right),\nonumber\\
    &&\vdots\label{E74}
\end{eqnarray}
can be constructed analogically from the undeformed counterpart. Corresponding time-derivatives,
\begin{eqnarray}
    &&\frac{dQ^1_d}{dt}=2\int\,dx\,\text{Im}\left(q_d^*(x,t){\cal  Y}(-x,t)\right),\nonumber\\
    &&\frac{dQ^2_d}{dt}=4\int\,dx\,\text{Re}\left(q_{d,x}^*(x,t){\cal  Y}(-x,t)\right),\nonumber\\
    &&\frac{dQ^3_d}{dt}=2\int\,dx\,\text{Re}\left(q_{d,t}^*(-x,t){\cal  Y}^*(x,t)\right),\nonumber\\
    \vdots\label{E75}
\end{eqnarray}
have ${\cal PT}$-odd integrands as required. It is not difficult to see that the general quasi-conservation argument made for NLS charges will also apply here, as apart from the non-locality, their analytical structure is the same. 

The particular 1-soliton solution for the non-local NLS system is known as \cite{R23},
\begin{equation}
    q^{\text{1-Sol}}_x(x,t)=-2\left(\eta_1+\eta_2\right)\frac{\exp\left[i\left(\theta_1-4\eta_2^2t\right)-\eta_2x\right]}{1+\exp\left[i\left\{\theta_1+\theta_2+4\left(\eta_1^2-\eta_2^2\right)t\right\}-2\left(\eta_1+\eta_2\right)x\right]},\label{NEq4}
\end{equation}
with free parameters $\eta_{1,2},\theta_{1,2}$, and does not possess a definite ${\cal PT}$-behavior. Other solutions have also been obtained recently \cite{R029,R030} for this system. However, particular quasi-deformation of this system is yet to be attempted, and it is beyond the scope of the present work. Soliton solutions have also been obtained for a non-local AB KdV system \cite{R031} that is applicable to many natural and social corelation phenomena. A particular multi-soliton of this system has the form \cite{R031},
\begin{equation}
    u(x,t)=2\left[\ln\sum_\nu K_\nu \cosh\left(\frac{1}{2}\sum_{j=1}^N\nu_j\eta_j\right)\right]_{xx},\quad\eta_j=k_jx-k_j^3t,\label{NEq5}
\end{equation}
with $\nu=\left\{\nu_1,\cdots,\nu_N\right\}$, $\nu_i=\pm 1$ and wave-numbers $k_j$ which is ${\cal PT}$-even. However, it remains to see if this property prevails over the quasi-deformation of this system. We expect that particular quasi-deformations of such non-local systems in the near future will also support the proposed framework.

\subsection{The Abelianization process and ${\cal PT}$-symmetry}
The quasi-conserved charges can be constructed formally using the usual Abelianization approach \cite{R1}. For the KdV system, it is achieved through the $sl(2)$ loop algebra,
\begin{eqnarray}
    && \left[F^m,\,F_\pm^n\right]=2F_\mp^{m+n},\quad\left[F_-^m,\,F_+^n\right]=F^{m+n+1};\nonumber\\
    && F^n=\lambda^n\sigma_3,\quad F_\pm^n=\frac{\lambda^n}{\sqrt{2}}\left(\sigma_+\pm\lambda\sigma_-\right),\label{E57}
\end{eqnarray}
with the spectral parameter $\lambda\in\mathbb{R}$, constructed from the inherent $su(2)$ algebra. Under ${\cal PT}$-transformations:
\begin{equation}
    {\cal PT}F^n\left({\cal PT}\right)^{-1}=F^n,\quad{\cal PT}F_\pm^n\left({\cal PT}\right)^{-1}=-F_\pm^n,\label{E58}
\end{equation}
this loop algebra remains unchanged. The ${\cal PT}$-odd quasi-deformed Lax pair is then gauge-rotated,
\begin{equation}
L^d_\mu\to\Tilde{L}^d_\mu=gL^d_\mu g^{-1}+\left(\partial_\mu g\right)g^{-1},\label{63}
\end{equation}
by the gauge operator $g=e^G,\quad G=\sum_{n=-1}^{-\infty}\left(\alpha_nF_-^n+\beta_nF^n\right)$ so that the spatial component $\Tilde{A}_d$ is now exclusively in the image of $sl(2)$. From the expressions of the coefficients $\alpha_n,\beta_n$ in Ref. \cite{R7}, 
\begin{equation}
    {\cal PT}\alpha_n\left({\cal PT}\right)^{-1}=-\alpha_n,\quad{\cal PT}\beta_n\left({\cal PT}\right)^{-1}=\beta_n.\label{NEq6}
\end{equation}
From these transformation properties, along with those in Eq. \ref{E58}, $G$ and thereby the gauge operator $g$ turns out as ${\cal PT}$-even:
\begin{equation}
    {\cal PT}G\left({\cal PT}\right)^{-1}=G\Rightarrow{\cal PT}g\left({\cal PT}\right)^{-1}=g,\label{E60}
\end{equation}
thereby maintaining ${\cal PT}$-oddness of the gauge-rotated Lax pair ${\cal PT}\Tilde{L}^d_\mu\left({\cal PT}\right)^{-1}=-\Tilde{L}^d_\mu$.  

\paragraph*{}Since the abelianization process maintains the ${\cal PT}$-structure of the quasi-KdV system, the corresponding quasi-conservation should again be due to a definite ${\cal PT}$-behavior. Indeed, as $F_{tx}\to gF_{tx}g^{-1}$, the quasi-conserved charges are obtained as \cite{R7},
\begin{equation}
    \frac{d}{dt}Q^n(t)=\int\,dx\,f_n^+{\cal X}=\Gamma^n.\label{E65}
\end{equation}
where $f^+_n=f^+_n\left(u_d\right)$ are given in Ref. \cite{R7}. Since ${\cal Y}^{PT}=-{\cal Y}\Rightarrow{\cal X}^{PT}={\cal X}$ and $\left(f_n^+\right)^{PT}=-f_n^+$ in the symmetric phase, the integrand is ${\cal PT}$-odd as required.

\section{Discussions and Conclusions}\label{Se3}
We have seen that quasi-conservation of a deformed integrable system can be assured by ${\cal PT}$-symmetry, given that the system is in the symmetric phase ($u_d^{PT}=\pm u_d$). Consequently, both the anomaly function ${\cal Y}$ and, thereby, the anomalous charges are bound to have definite ${\cal PT}$ properties required for quasi-conservation \cite{R1,R3,R7}. This system is characterized by a ${\cal PT}$-odd Lax pair responsible for a geometric phase of evolution that mimics the condition for integrability. This structure is expected to prevail for any deformation that does not violate the original ${\cal PT}$-symmetry of the system, such as abelianization\footnote{The abelianization is expected to be so, as it is based on the inherent loop algebra of the system.} or otherwise. It should be noted that integrability does not explicitly require ${\cal PT}$-symmetry; the corresponding charges are conserved locally following the equation of motion. But for quasi-integrable models, asymptotic conservation follows directly from ${\cal PT}$-symmetry in the absence of any zero-curvature condition. Indeed, it can be possible that the ${\cal PT}$-symmetry of a system can ensure asymptotic conservation of charges, whether the system is integrable or not.

\paragraph*{} Demanding ${\cal PT}$-symmetry imposes a strong constraint on the particular deformation in order to cause quasi-integrability. This is true for the KdV \cite{R7} and NLS \cite{R3} systems and is expected to be true for other systems. As for particular quasi-deformed solutions obtained for various systems \cite{R1,R3,R6,R8,R9,R21}, they always display definite-${\cal PT}$ structures in the form of single- and multi-soliton-like structures that are fairly stable. Although sufficient, note that ${\cal PT}$-symmetry is not hailed here as the necessary condition for quasi-integrability. However, since definite ${\cal PT}$-property of anomalous charges is essential for quasi-conservation, it is hard to see if that can be obtained without definite-${\cal PT}$ anomaly and solution.

\paragraph*{}It would be interesting to look for explicit examples of quasi-integrability in ${\cal PT}$-symmetric nonlinear models \cite{R18,R19,R20}. Indeed, nonlinearity is seen to `repair' the broken-${\cal PT}$ phase \cite{R18,R19}. This could imply non-trivial symmetric phases that can support quasi-conservation. Most of such systems are also non-Hermitian and yet there are localized solutions, usually in terms of optical excitations. It can be possible that even if the Hermitian counterpart was not integrable its ${\cal PT}$-symmetric analog becomes one, although under the present formulation the latter needs to be a quasi-deformation of an integrable system.

\section*{Acknowledgements}
Kumar Abhinav’s research was funded by Mahidol University (Fundamental Fund: fiscal year 2025 by the National Science Research and Innovation Fund (NSRF)) and he deeply acknowledges many enlightening discussions with Professor Prasanta K. Panigrahi. Partha Guha is grateful to Professors Luiz A. Ferreira and Wojtek J. Zakrzewski for various useful discussions.

\section*{Funding statement}
Kumar Abhinav’s research was funded by Mahidol University (Fundamental Fund: fiscal year 2025 by the National Science Research and Innovation Fund (NSRF)).

\section*{Author contributions statement}
K. A. co-conceived the idea, did the initial calculations, performed the analysis, wrote and communicated the manuscript. P. G. and I. M. co-conceived the idea, supervised the progress of the manuscript and provided explanations for the key concepts. All authors reviewed the manuscript.

\section*{Competing interests}
The authors declare no competing interests.

 \section*{Data availability}
All data that support the findings in this study are available in the article. Additional information is available
from the corresponding author upon request.

\end{document}